\def\flash{1}
\def\aa{2}
\def\iit{3}
\def\cfa{4}
\def\umass{5}

\newcommand{\iso}[2]{$^{#2}${#1}}

\newcommand{\msolar}{\ensuremath{M_{\odot}}}

\newcommand{\kms}{km s$^{-1}$}

\newcommand{\sneia}{SNe~Ia}

\newcommand{\gcjivb}{J12}

\newcommand{\rhoc}{\ensuremath{\rho_{c}}}

\documentclass[apjl]{emulateapj}

\begin{document}

\title{Failed-Detonation Supernovae: Sub-Luminous Low-Velocity Ia Supernovae \\
and Their Kicked Remnant White Dwarfs with Iron-Rich Cores}

\author{
George~C.~Jordan IV,\altaffilmark{\flash,\aa}
Hagai~B.~Perets,\altaffilmark{\iit,\cfa}
Robert~T.~Fisher\altaffilmark{\umass}
Daniel~R.~van~Rossum,\altaffilmark{\flash,\aa}
}
\altaffiltext{\flash}{Center for Astrophysical Thermonuclear Flashes, The
University of Chicago, Chicago, IL 60637.}

\altaffiltext{\aa}{Department of Astronomy and Astrophysics, The
University of Chicago, Chicago, IL 60637.}

\altaffiltext{\iit}{Physics department, Technion - Israel Institute of Technology,
Haifa, Israel 32000.}

\altaffiltext{\cfa}{Harvard-Smithsonian Center for Astrophysics, 60 Garden St., Cambridge, USA 02138}

\altaffiltext{\umass}{University of Massachusetts Dartmouth, Department
of Physics, 285 Old Westport Road, North Dartmouth, 02740.}

\begin{abstract}
Type Ia supernovae (\sneia) originate from the thermonuclear
explosions of carbon-oxygen (C-O) white dwarfs (WDs).
The single-degenerate scenario is a well-explored
model of \sneia\ where unstable thermonuclear burning initiates in an
accreting, Chandrasekhar-mass WD and forms an advancing flame.
By several proposed physical processes the rising, burning material
triggers a detonation, which subsequently consumes and unbinds the WD.
However, if a detonation is not triggered and the deflagration is too weak
to unbind the star, a completely different scenario unfolds.
We explore the failure of the Gravitationally-Confined Detonation (GCD) mechanism
of \sneia, and demonstrate through 2D and 3D simulations the properties of failed-detonation SNe.
We show that failed-detonation SNe expel a few 0.1 \msolar\ of burned
and partially-burned material
and that a fraction of the material
falls back onto the WD, polluting the remnant WD with intermediate-mass
and iron-group elements, that likely segregate to the core forming
an WD whose core is iron rich.
The remaining material is asymmetrically ejected
at velocities comparable to the escape velocity from the WD, and in
response, the WD is kicked to velocities of a few hundred km s$^{-1}$.
These kicks may unbind the binary and eject a runaway/hyper-velocity
WD.
Although the energy and ejected mass of the failed-detonation
SN are a fraction of typical thermonuclear
SNe, they are likely to appear as sub-luminous
low-velocity \sneia.
Such failed detonations might therefore
explain or are related to the observed branch of peculiar \sneia,
such as the family of low-velocity sub-luminous SNe (SN 2002cx/SN
2008ha-like SNe).
\end{abstract}


\section{Introduction}
Type Ia supernovae (\sneia) are among the most energetic explosions in the known universe,
releasing $\sim 10^{51}$ ergs of kinetic energy in their ejecta, and synthesizing $\sim 0.7$ \msolar\ of 
radioactive \iso{Ni}{56}.
The discovery of the Phillips relation \citep{pskovskii77, phillips93} enabled the use of \sneia\ as standardizable
cosmological candles, and has ushered in a new era of astronomy leading to the discovery of the
acceleration of the universe \citep{1998AJ....116.1009R, schmidtetal98, 1999ApJ...517..565P}, and to the 2011 Nobel Prize in physics.

Models of normal \sneia, such as the single degenerate (SD) model, focus on exploding the WD in order to produce the
explosion energies, luminosities, and typical velocities observed in normal \sneia.
This is accomplished either by consuming enough of the WD with the initial subsonic buring phase --- or deflagration phase --- to unbind the WD
as theorized by the Pure Deflagration (PD) \citep{2005ApJ...623..337G, 2005A&A...431..635R} model,
or consuming the entire WD by a detonation triggered by the deflagration phase
as posited by the deflagration-to-detonation transition
(DDT) \citep{1991A&A...245..114K, 2004PhRvL..92u1102G, 2005ApJ...623..337G},
the Pulsating Reverse Detonation (PRD) \citep{2009ApJ...695.1244B, 2009ApJ...695.1257B},
and the GCD \citep{2004astro.ph..5162C,2008ApJ...681.1448J, 2009ApJ...693.1188M, 2012ApJ...759...53J} models of \sneia.

In the following, we present a novel variant of
SD model of \sneia\ in which the
deflagration is too weak to unbind the star\footnote{We acknowledge related
work by \citet{2012arXiv1210.5243K} that appeared as this article was going to press.} and
a detonation is not triggered by any of the proposed mechansisms,
resulting in the survival of a bound remnant of the original WD.
We present for the first time predictions of these failed-detonation (FD) SNe from 2D and 3D
simulations.
We show that FD models have numerous remarkable implications for the observable properties of the
resulting explosion and its outcomes.
These include the production of a family of peculiar \sneia\ events with low expansion velocities, low luminosities and low ejecta-mass ---
whose properties are
broadly consistent with the observed properties of a branch of peculiar \sneia\ similar to SN 2002cx and/or SN 2008ha.
Even more remarkably, the remnant WD receives a large velocity kick
from the asymmetric nature of the deflagration, and is enriched with both intermediate-mass (IME's)
and iron-group elements (IGE's),
forming a peculiar WD with a heavey/iron-rich core.

Previous work has suggested that though the PD model has shortcomings explaining normal \sneia,
they may explain 2002cx-like SNe \citep{branchetal04}.
The WD, however, is fully incinerated in these models, producing a Chandrasekhar-mass of ejecta.
Such models might therefore not be able to explain the large diversity recently observed among
SNe of this peculiar class of SNe.
Additionally, a study by \citet{livneetal05} of initial conditions for the PD model produced
situations that they termed ``fizzles'' which did not produce a healthy PD explosion and left
the WD bound.
They did not pursue these models beyond the scope of their study,
nor did they relate these fizzles to peculiar \sneia.

\section{Avoiding the Transition to Detonation\label{SEC:PHYSS}}
The FD scenario requires that no detonation is triggered as a result of the deflagration event.
We briefly touch on the possibility of the PRD and GCD failing to trigger a detonation. 
We first make the general assumption that is made in the PD, PRD, and GCD scenarios, namely that
the DDT mechanism is not active.
We refer the reader to \citet{2009ApJ...695.1244B} for a discussion of the justification of
this assumption, the details of which are beyond the scope of this work.

In the PRD model, ash ejected during the deflagration phase falls back onto the WD.
An accretion shock formed by infalling ash surrounds and heats fuel remaining
in the WD core, which in turn triggers a detonation.
\citet{2009ApJ...695.1244B} reported that the PRD could not trigger a detonation
if the energy released during the deflagration was near that of the binding energy
of the WD. This situation is realized in our simulations.

\citet{2012ApJ...759...53J} (hereafter \gcjivb) detail how the GCD mechanism
triggers a gradient-induced detonation when ash from the deflagration
flows over the stellar surface, mixes with cold fuel, collides at the
antipodal point from break out, and is squeezed to the
necessary temperatures and densities by the contracting WD.
Models presented here show that a detonation via the GCD mechanism
is avoided alltogether.
The primary difference between these models and those of \gcjivb\ is the amount of energy released during the
deflagration.
More energy is released and delivered to the WD than in \gcjivb;
thus, more mass is consumed by the flame and ejected from the WD than previously.
The WD is modified to a higher degree and as a result
the WD can not contract enough to squeeze the fuel-ash mixture to the critical conditions for detonation.
Therefore the star never detonates.
We note that \citet{2007ApJ...660.1344R} also investigated the failure of the GCD mechanism;
however, their simulations in which the WD was still bound and
star did not detonate were stopped before the contraction phase.
\gcjivb\ discussed these models and showed that
had \citet{2007ApJ...660.1344R} run their simulations
longer, they would have most likely triggered a GCD on contraction. 

The FD scenario thus occurs when a weak deflagration leaves a partially bound WD (in constrast to the PD model)
and the conditions for detonation are never realized (and the DDT is avoided).

\section{Simulations of the FD Model\label{sec:FD-models}}
\subsection{Simulation Setup\label{subsec:simsetup}}
We used the Adaptive Mesh Refinement (AMR) FLASH application framework \citep{dubey2009, 2000ApJS..131..273F} to perform our
simulations of FD models.
FLASH has been previously used to simulate \sneia\ in both 2D cylindrical
and 3D Cartesian geometry (\cite{2008ApJ...681.1448J, 2009ApJ...693.1188M, 2012ApJ...757..175K}, \gcjivb).
Our simulations include an advection-diffusion-reaction (ADR) treatment of the thermonuclear
flame \citep{2007ApJ...656..313C, 2007ApJ...668.1118T, 2009ADNDT..95...96S},
an equation of state that includes contributions from blackbody radiation, ions,
and electrons of an arbitrary degree of degeneracy \citep{2000ApJS..126..501T},
and the multipole treatment of gravity \citep{ASC2012}.

We performed four 3-dimensional (3D) exploratory simulations to test the feasibility of the FD
and 1 full 2-dimensional (2D) model to adequately observe the FD at late times.

Our 3D simulations included a reduced domain size and a moderate resolution
(8 km) which reduced their computational expense.
We initialized these simulations similarly to those in \gcjivb\
with a 1.365 \msolar\ WD placed at the origin of the domain.
We used the same distribution of sixty three, $16$ km radius ignition ``points''
distributed in a $128$ km radius spherical volume.
We chose 48 km, 38, km, 28 km, and 18 km as offset distances along the z-axis of the spherical volume.
Note that these offsets are closer to the WD core than in \gcjivb\ and lead
to more buring during the deflagration phase.
Table \ref{tab:sims} lists the initial conditions for each simulation and the corresponding names
we gave them.
We ran the simulations from ignition, through peak
stellar expansion, and at least until the WD reached a maximum central density upon contraction.

Our 2D simulation was performed with a large domain in 2D cylindrical geometry
at 4 km resolution and ran for 60 seconds.
This simulation was initialized similarly to the 3D simulations except we placed only 4 bubbles
in a 64 km spherical volume offset by 70 km along the z-axis (axis of symmetry).
We chose these initial conditions to obtain an FD in 2D given what we learned from
our 3D simulations.
The larger domain size allowed us to follow the outer layers of the ejecta for the
entirety of the simulation.

\subsection{Simulation Results\label{subsec:simresults}}

We obtained an FD from each of our simulations.
Table \ref{tab:sims} contains a collection of their properties.

\begin{figure}
\begin{center}
\includegraphics[width=0.5 \textwidth]{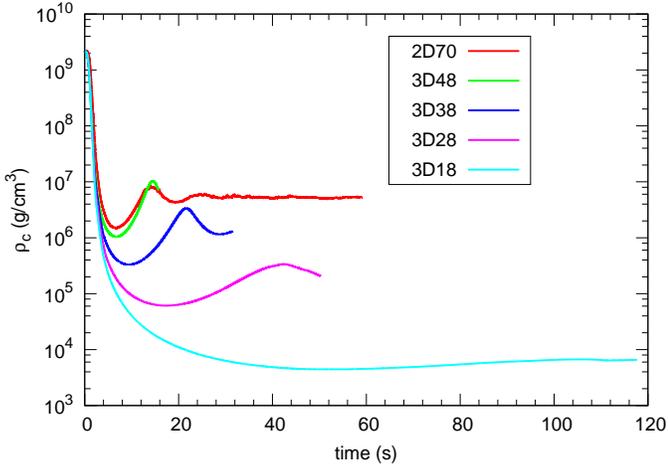}
\caption{
The central density, \rhoc,  of the WD vs time.
\label{fig:rhocvstime}}
\end{center}
\end{figure}

The deflagration liberated between 89\% and
167\% of the binding energy of the WD.
However, in each simulation the deflagration fails to unbind the WD.
The WD expands in response to the deflagration, reaches its maximum
level of expansion, and then contracts.
Figure \ref{fig:rhocvstime} shows the evolution of the central density,
\rhoc, of the WD and illustrates the oscillatory nature of the WD after the
deflagration.
The more energy released during the deflagration, the
more the star expanded, the longer
the pulsational period, and the smaller \rhoc\ at maximum contraction.

The fact that the deflagration releases more than 100\%
WD binding energy in some models suggests that the WD should be unbound after the deflagration;
however, the entire energy budget does not work to only unbind the star.
For example, some of the energy is lost when the ejecta and the remnant WD are accelerated to high velocities.
Even though the deflagration liberates enough energy to completely unbind
the star, the energy is partitioned in such a way that a portion of the original WD
remains gravitationally bound.

In each of our models, the WD gets a kick in the opposite direction from which the buoyant ash rises and breaks
through the surface of the star.
We measured this velocity to be on the order of hundreds of \kms and 
list these values for each simulation in table \ref{tab:sims}.
We note that momentum conservation is better than 10\% of the momentum of the
WD kick in our simulations.
Some of the material from the FD escapes and achieves high velocities, while some of
the material is bound to the remnant WD and will eventually accrete onto its surface.
Table \ref{tab:sims} lists the composition of material that remains gravitationally
bound and eventually will mix with the remnant WD.
In all cases between 0.04 \msolar\ and 0.2 \msolar\ of IME's and IGE's (some of which is radioactive
\iso{Ni}{56}) remained bound to the star.
The evolution of the WD as the material falls back onto and heats
the star is an interesting question and one which we will examine in future work.

The composition of the material that escapes the system is also listed
in table \ref{tab:sims} along with the kinetic energy and the
mass-weighted velocity of the ejecta.
This material includes carbon and oxygen, IME's, and IGE's, and
ranges from 0.2 to 1.0 \msolar.
In general, the more material that is burned during the deflagration,
the more material that escapes.

\begin{figure*}[th!]
\begin{center}
\includegraphics[width=1.0 \textwidth]{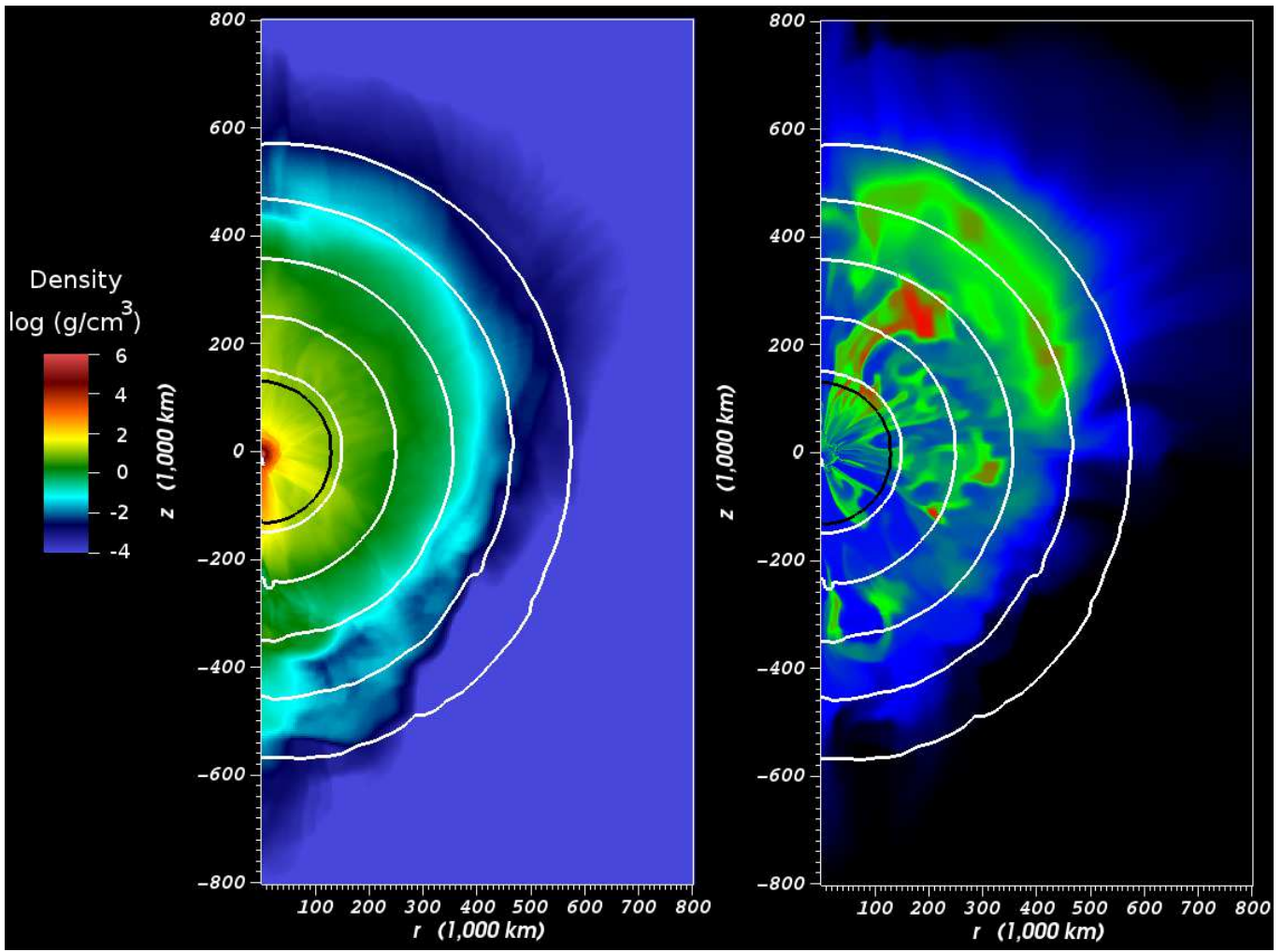}
\caption{
Images from the $2D70$ simulation.
The black contour marks the transition between
gravitationally bound and escaping material.
The white contours are of the magnitude of
the velocity field.
From the inner most contour and moving outwards, the
values are: 2,000 km s$^{-1}$, 4,000 km s$^{-1}$,
6,000 km s$^{-1}$, 8,000 km s$^{-1}$, and 10,000 km s$^{-1}$.
The images show the simulation at 60 seconds.
$(left)$ Log density of the remnant.
The values of the density are given by the color bar on the
left.
$(right)$ The composition of the remnant.
Blue is C-O, green is IME's, and
red is IGE's.
\label{fig:combo}}
\end{center}
\end{figure*}

Figure \ref{fig:combo} shows
shows the density structure and composition with
overlaid velocity contours of the 2D model at 60s.
Note that the density profile of the FD model is asymmetric in velocity space
between the hemisphere corresponding to the ejected deflagration
and the opposite hemisphere of the system.
The remnant of the WD can be seen as the tiny high-density feature slightly below the
origin of the domain.

The figure also shows the nature of the asymmetry in composition of the structure.
The north side of the remnant contains the products of the deflagration that were
sprayed from the surface of the WD.
Clumpy structures of IME's and IGE's exist at a range of velocities in the
northern hemisphere of the domain but are less abundant in the southern hemisphere.
The asymmetries suggest that this object would look much different depending on the viewing
angle of the observer.

All of our models produce a relatively small amount of radioactive \iso{Ni}{56}.
Though we do not perform detailed nucleosynthetic post processing of the 3D models,
we can set an upper limit on the \iso{Ni}{56}\ yields with the amount of IGE's produced,
which ranges from 0.2 \msolar\ to 0.34 \msolar.
Neutronization through electron capture reactions during the deflagration
would shift production away from \iso{Ni}{56} though and reduce its contribution
to the IGE totals.
Whether even lower masses of IGE's (and thus \iso{Ni}{56}) could be produced
under appropriate conditions (e.g. comparable to that observed in SN 2008ha, a very faint SN with extremely
low ejecta velocities), is yet to be explored.

In summary, we find that a remnant of the WD survives the FD SN event with a lower mass than the original.
The remaining bound stellar material is kicked by the ejection of the ash
and obtains a velocity of hundreds of km s$^{-1}$.
An asymmetric outburst of deflagration products
rich in IGE's (such as Fe, Co, and Ni) and containing some
IME's (such as Mg, Si, and S) is produced.
Some of this material attains escape velocity and some falls
back onto the star.
The velocity of the escapting outflow was slow (approximately a few thousand km s$^{-1}$) related
to normal \sneia\ as a result of the comparatively small amount of energy released in the FD scenario.
Due to the weak nature of the deflagration, the FD only converted 15\% and 25\% of the WD
to \iso{Ni}{56}, compared to normal \sneia\ which convert nearly half of the WD to \iso{Ni}{56}.
Finally, the collection of simulations we present produces a range of values for each of the discussed
properties which suggests a population of events within this class of model.

\section{Discussion and Predictions\label{sec:discussion}}
\subsection{Sub-luminous Low-Velocity SNe}

The most prominent features of the FD models are the low mass and low velocity of the ejecta,
which translate into the production of typically
sub-luminous, low-velocity \sneia.
It is therefore natural to examine whether SNe with such characteristics have
already been discovered.
In particular, one may explore peculiar SNe exhibiting either extremely low mass ejecta, such as SN 2008ha \citep{fol+09,val+09},
or low ejecta velocity such as SN 2002cx-like SNe \citep{lietal03, branchetal04, jhaetal06}.

Normal \sneia\ differ in their ejecta velocities as measured in
some standard method, but they generally fall between $9000-14000$
km s$^{-1}$ near peak luminosity, with similar dispersion at later
times (as derived from the Si II line \citep{ben+05}).
The velocities of even the lowest velocity
SNe in the \citet{ben+05} sample much exceed the mass-averaged FD velocities 
listed in table \ref{tab:sims}.
We also note a trend in our simulations of more energetic and likely more luminous (larger IGE yield) SNe to be
accompanied by higher ejecta-velocities over almost an order of magnitude in kinetic energy.  
The only other type of SNe with such low
expansion velocities are the branch of peculiar type Ia SNe, named
for the prototype for this class of supernovae, 2002cx \citep{lietal03, branchetal04, jhaetal06};
such SNe may also have an energy-velocity correlation \citep{mcc+10} as observed in our simulations.   

SN 2002cx-like events are characterized by luminosities which lie
too low in comparison to the Phillips relation for Branch-normal Ia
events \citep{lietal03}, low photospheric velocities \citep{lietal03},
weak intermediate-mass element lines
\citep{branchetal04}, and late-time optical nebular spectra
dominated by narrow Fe II lines \citep{branchetal04, jhaetal06}.
Since the discovery of SN 2002cx, a number of other 2002cx-like events
have been discovered, including 2002es, 2005P, 2005hk \citep{chornocketal06},
2008ge \citep{foleyetal10} and 2008ha \citep{fol+09,val+09}. The latter event (SN 2008ha),
in particular, is consistent with extremely low mass ejecta and energetics.
We predict the FD models to produce similar properties to those characterizing SN 2002cx like SNe,  given the low expansion
velocities and the low estimated \iso{Ni}{56} yield, and potentially even explaining SN 2008ha like events with low mass ejecta.

Though our initial set of simulations is limited,
the robust features of FD's, including low velocity ejecta, the expected low luminosity (due to the small yield of \iso{Ni}{56})
and their low mass ejecta
(comparable to that SN 2008ha) make them tantalizing candidate
progenitors for this branch of peculiar SNe.
Note that the
single degenerate origin of such SNe is also consistent with the overall
typically young (but not \emph{necessarily} young; \citeauthor{fol+10} 2010) environments
found for SN-2002cx like SNe, compared to the expectations from, e.g.
core-collapse SNe (only very young environments).

\subsection{WD's with Heavy/Iron-Rich Cores\label{subsec:fecores}}

In our FD scenario, a large amount of burnt material falls back
to the remnant WD.
From table \ref{tab:sims}, the WD may incorporate as much
as $0.3$ $M_{\odot}$ of IGE's and $0.07$ $M_{\odot}$ of IME's
of fallback material, together comprising
as much as $\sim 18\%$ of the remnant C-O WD.
In time, these elements are likely to gravitationally settle to the WD core,
creating WD's with iron-rich/heavy core.
The existence of iron-core
WD's has been considered before, with even the potential observation
of such WDs \citep{pro+98,cat+08}.
The FD scenario therefore 
provides a novel evolutionary scenario for the formation of these iron/heavy-core
C-O WD's. A somewhat related scenario of failed SN was suggested for
the formation of O-Ne-Mg WDs with iron cores \citep{ise+91}.

\subsection{WD Natal Kicks}

FD's produce a highly asymmetric ejection of material.
This is not unique amongst various models for SNe explosions.
However, in our FD case, the WD survives the explosion.
Considering momentum conservation, this gives rise to a unique outcome,
namely that the surviving WD is kicked at very high velocities, ranging hundreds of
km s$^{-1}$.
The FD scenario suggests
the existence of strong WD natal kicks,
and provides an
interesting prediction per the existence of hypervelocity WDs. Taken
together, the potential existence of a a heavy core WD (discussed in section \ref{subsec:fecores}),
and the high ejection velocity produce a highly peculiar
object, which, if observed may provide a possibly unique smoking gun
signature.
One should note, however, that the population of halo WDs
may also have relatively high velocities, and it might therefore be
difficult to pinpoint the kinematic property as related to a natal
kick (unless the WD is massive and young; an unlikely possibility
for the old population of halo WDs).

We note that velocities of hundreds of km s$^{-1}$ could be larger
than the orbital velocities of the SN binary progenitor, and a kick velocity
of such magnitude can therefore unbind the binary.
Various binary configurations have been explored for the single degenerate progenitor
models, including progenitors with MS and RG companions \citep{1996ApJ...470L..97H, 2000ApJS..128..615M}.
We conclude that the range of WD kick velocities could either unbind the
binary (more likely for WD-MS binaries), or leave behind a bound WD binary (more likely for the WD-RG
binaries). The latter case could
lead to the formation of a very compact, but potentially eccentric WD-binary, which would be difficult to
produce through other channels of binary evolution.

\section{Summary}
Our simulations demonstrate the properties of a scenario in which the deflagration is too
weak to unbind the WD and the conditions to trigger a detonation are not met.
These failed-detonations result in an asymmetric outburst of deflagration material
consisting of IME's and IGE's along with a fraction of the original
WD still gravitationally bound.
The models produce a family of faint \sneia\ with a slowly evolving light curves due to the low \iso{Ni}{56}\ yield
and the low energetics.
The remaining WD gets a kick on the order of hundreds of km s$^{-1}$ and is
contaminated with fall-back from the deflagration, producing a WD with an iron-rich/heavy core.
We presented our initial simulations to quantify the some of the bulk observable properties and demonstrate the
conditions under which the GCD fails.
We further hypothesize that the FD model is a possible explanation for 2002cx-like SN.
Future studies will explore the detailed observational features of FD SNe and their direct comparison to observations.



\acknowledgements
The authors thank the FLASH Code Group, the FLASH Astrophysics Group,
and the Argonne Leadership Computing
Facility at Argonne National Laboratory.
HBP is supported by the CfA and BIKURA prize fellowships.
This work was supported in part at the University of Chicago
by the U.S Department of Energy (DOE) under Contract B523820 to the ASC
Alliances Center for Astrophysical Nuclear Flashes, and in part by the
National Science Foundation under Grant No. AST - 0909132
for the ``Petascale Computing of Thermonuclear Supernova Explosions''.
This research used computational resources awarded under the INCITE program
at ALCF at ANL, which is supported by the Office of Science of the US
Department of Energy under Contract No. DE-AC02-06CH11357.


\begin{thebibliography}{50}
\expandafter\ifx\csname natexlab\endcsname\relax\def\natexlab#1{#1}\fi

\bibitem[{{Arnett} \& {Livne}(1994)}]{1994ApJ...427..330A}
{Arnett}, D. \& {Livne}, E. 1994, \apj, 427, 330

\bibitem[{{ASC FLASH Center}(2012)}]{ASC2012}
{ASC FLASH Center}. 2012, FLASH User's Guide, 4th edn., University of Chicago,
  http://flash.uchicago.edu/website/codesupport/flash4\_ug/

\bibitem[{{Benetti} {et~al.}(2005)}]{ben+05}
{Benetti}, S. {et~al.} 2005, \apj, 623, 1011

\bibitem[{{Branch} {et~al.}(2004){Branch}, {Baron}, {Thomas}, {Kasen}, {Li}, \&
  {Filippenko}}]{branchetal04}
{Branch}, D., {Baron}, E., {Thomas}, R.~C., {Kasen}, D., {Li}, W., \&
  {Filippenko}, A.~V. 2004, \pasp, 116, 903

\bibitem[{{Bravo} \& {Garc{\'{\i}}a-Senz}(2009)}]{2009ApJ...695.1244B}
{Bravo}, E. \& {Garc{\'{\i}}a-Senz}, D. 2009, \apj, 695, 1244

\bibitem[{{Bravo} {et~al.}(2009){Bravo}, {Garc{\'{\i}}a-Senz}, {Cabez{\'o}n},
  \& {Dom{\'{\i}}nguez}}]{2009ApJ...695.1257B}
{Bravo}, E., {Garc{\'{\i}}a-Senz}, D., {Cabez{\'o}n}, R.~M., \&
  {Dom{\'{\i}}nguez}, I. 2009, \apj, 695, 1257

\bibitem[{{Calder} {et~al.}(2004){Calder}, {Plewa}, {Vladimirova}, {Lamb}, \&
  {Truran}}]{2004astro.ph..5162C}
{Calder}, A.~C., {Plewa}, T., {Vladimirova}, N., {Lamb}, D.~Q., \& {Truran},
  J.~W. 2004, ArXiv Astrophysics e-prints

\bibitem[{{Calder} {et~al.}(2007){Calder}, {Townsley}, {Seitenzahl}, {Peng},
  {Messer}, {Vladimirova}, {Brown}, {Truran}, \& {Lamb}}]{2007ApJ...656..313C}
{Calder}, A.~C., {Townsley}, D.~M., {Seitenzahl}, I.~R., {Peng}, F., {Messer},
  O.~E.~B., {Vladimirova}, N., {Brown}, E.~F., {Truran}, J.~W., \& {Lamb},
  D.~Q. 2007, \apj, 656, 313

\bibitem[{{Catal{\'a}n} {et~al.}(2008){Catal{\'a}n}, {Ribas}, {Isern}, \&
  {Garc{\'{\i}}a-Berro}}]{cat+08}
{Catal{\'a}n}, S., {Ribas}, I., {Isern}, J., \& {Garc{\'{\i}}a-Berro}, E. 2008,
  \aap, 477, 901

\bibitem[{{Chornock} {et~al.}(2006){Chornock}, {Filippenko}, {Branch}, {Foley},
  {Jha}, \& {Li}}]{chornocketal06}
{Chornock}, R., {Filippenko}, A.~V., {Branch}, D., {Foley}, R.~J., {Jha}, S.,
  \& {Li}, W. 2006, \pasp, 118, 722

\bibitem[{Dubey {et~al.}(2009)Dubey, Antypas, Ganapathy, Reid, Riley, Sheeler,
  Siegel, \& Weide}]{dubey2009}
Dubey, A., Antypas, K., Ganapathy, M., Reid, L., Riley, K., Sheeler, D.,
  Siegel, A., \& Weide, K. 2009, Parallel Computing, 35, 512

\bibitem[{{Foley} {et~al.}(2009){Foley}, {Chornock}, {Filippenko},
  {Ganeshalingam}, {Kirshner}, {Li}, {Cenko}, {Challis}, {Friedman}, {Modjaz},
  \& {Wood-Vasey}}]{fol+09}
{Foley}, R.~J., {Chornock}, R., {Filippenko}, A.~V., {Ganeshalingam}, M.,
  {Kirshner}, R.~P., {Li}, W., {Cenko}, S.~B., {Challis}, P., {Friedman},
  A.~S., {Modjaz}, M., \& {Wood-Vasey}, W.~M. 2009, ArXiv:0902.2794

\bibitem[{{Foley} {et~al.}(2010{\natexlab{a}}){Foley}, {Rest}, {Stritzinger},
  {Pignata}, {Anderson}, {Hamuy}, {Morrell}, {Phillips}, \&
  {Salgado}}]{foleyetal10}
{Foley}, R.~J., {Rest}, A., {Stritzinger}, M., {Pignata}, G., {Anderson},
  J.~P., {Hamuy}, M., {Morrell}, N.~I., {Phillips}, M.~M., \& {Salgado}, F.
  2010{\natexlab{a}}, \aj, 140, 1321

\bibitem[{{Foley} {et~al.}(2010{\natexlab{b}}){Foley}, {Rest}, {Stritzinger},
  {Pignata}, {Anderson}, {Hamuy}, {Morrell}, {Phillips}, \& {Salgado}}]{fol+10}
---. 2010{\natexlab{b}}, \aj, 140, 1321

\bibitem[{{Fryxell} {et~al.}(2000){Fryxell}, {Olson}, {Ricker}, {Timmes},
  {Zingale}, {Lamb}, {MacNeice}, {Rosner}, {Truran}, \&
  {Tufo}}]{2000ApJS..131..273F}
{Fryxell}, B., {Olson}, K., {Ricker}, P., {Timmes}, F.~X., {Zingale}, M.,
  {Lamb}, D.~Q., {MacNeice}, P., {Rosner}, R., {Truran}, J.~W., \& {Tufo}, H.
  2000, \apjs, 131, 273

\bibitem[{{Gamezo} {et~al.}(2004){Gamezo}, {Khokhlov}, \&
  {Oran}}]{2004PhRvL..92u1102G}
{Gamezo}, V.~N., {Khokhlov}, A.~M., \& {Oran}, E.~S. 2004, Physical Review
  Letters, 92, 211102

\bibitem[{{Gamezo} {et~al.}(2005){Gamezo}, {Khokhlov}, \&
  {Oran}}]{2005ApJ...623..337G}
---. 2005, \apj, 623, 337

\bibitem[{{Hachisu} {et~al.}(1996){Hachisu}, {Kato}, \&
  {Nomoto}}]{1996ApJ...470L..97H}
{Hachisu}, I., {Kato}, M., \& {Nomoto}, K. 1996, \apjl, 470, L97

\bibitem[{{Isern} {et~al.}(1991){Isern}, {Canal}, \& {Labay}}]{ise+91}
{Isern}, J., {Canal}, R., \& {Labay}, J. 1991, \apjl, 372, L83

\bibitem[{{Jha} {et~al.}(2006){Jha}, {Branch}, {Chornock}, {Foley}, {Li},
  {Swift}, {Casebeer}, \& {Filippenko}}]{jhaetal06}
{Jha}, S., {Branch}, D., {Chornock}, R., {Foley}, R.~J., {Li}, W., {Swift},
  B.~J., {Casebeer}, D., \& {Filippenko}, A.~V. 2006, \aj, 132, 189

\bibitem[{{Jordan} {et~al.}(2008){Jordan}, {Fisher}, {Townsley}, {Calder},
  {Graziani}, {Asida}, {Lamb}, \& {Truran}}]{2008ApJ...681.1448J}
{Jordan}, IV, G.~C., {Fisher}, R.~T., {Townsley}, D.~M., {Calder}, A.~C.,
  {Graziani}, C., {Asida}, S., {Lamb}, D.~Q., \& {Truran}, J.~W. 2008, \apj,
  681, 1448

\bibitem[{{Jordan} {et~al.}(2012){Jordan}, {Graziani}, {Fisher}, {Townsley},
  {Meakin}, {Weide}, {Reid}, {Norris}, {Hudson}, \&
  {Lamb}}]{2012ApJ...759...53J}
{Jordan}, IV, G.~C., {Graziani}, C., {Fisher}, R.~T., {Townsley}, D.~M.,
  {Meakin}, C., {Weide}, K., {Reid}, L.~B., {Norris}, J., {Hudson}, R., \&
  {Lamb}, D.~Q. 2012, \apj, 759, 53



\bibitem[{{Khokhlov}(1991)}]{1991A&A...245..114K}
{Khokhlov}, A.~M. 1991, \aap, 245, 114

\bibitem[{{Khokhlov} {et~al.}(1997){Khokhlov}, {Oran}, \&
  {Wheeler}}]{1997ApJ...478..678K}
{Khokhlov}, A.~M., {Oran}, E.~S., \& {Wheeler}, J.~C. 1997, \apj, 478, 678

\bibitem[{{Kromer} {et~al.}(2012){Kromer}, {Fink}, {Stanishev}, {Taubenberger},
  {Ciaraldi-Schoolman}, {Pakmor}, {Roepke}, {Ruiter}, {Seitenzahl}, {Sim},
  {Blanc}, {Elias-Rosa}, \& {Hillebrandt}}]{2012arXiv1210.5243K}
{Kromer}, M., {Fink}, M., {Stanishev}, V., {Taubenberger}, S.,
  {Ciaraldi-Schoolman}, F., {Pakmor}, R., {Roepke}, F.~K., {Ruiter}, A.~J.,
  {Seitenzahl}, I.~R., {Sim}, S.~A., {Blanc}, G., {Elias-Rosa}, N., \&
  {Hillebrandt}, W. 2012, ArXiv e-prints

\bibitem[{{Krueger} {et~al.}(2012){Krueger}, {Jackson}, {Calder}, {Townsley},
  {Brown}, \& {Timmes}}]{2012ApJ...757..175K}
{Krueger}, B.~K., {Jackson}, A.~P., {Calder}, A.~C., {Townsley}, D.~M.,
  {Brown}, E.~F., \& {Timmes}, F.~X. 2012, \apj, 757, 175

\bibitem[{{Kuhlen} {et~al.}(2006){Kuhlen}, {Woosley}, \&
  {Glatzmaier}}]{kuhlenetal06}
{Kuhlen}, M., {Woosley}, S.~E., \& {Glatzmaier}, G.~A. 2006, \apj, 640, 407

\bibitem[{{Li} {et~al.}(2003){Li}, {Filippenko}, {Chornock}, {Berger},
  {Berlind}, {Calkins}, {Challis}, {Fassnacht}, {Jha}, {Kirshner}, {Matheson},
  {Sargent}, {Simcoe}, {Smith}, \& {Squires}}]{lietal03}
{Li}, W., {Filippenko}, A.~V., {Chornock}, R., {Berger}, E., {Berlind}, P.,
  {Calkins}, M.~L., {Challis}, P., {Fassnacht}, C., {Jha}, S., {Kirshner},
  R.~P., {Matheson}, T., {Sargent}, W.~L.~W., {Simcoe}, R.~A., {Smith}, G.~H.,
  \& {Squires}, G. 2003, \pasp, 115, 453

\bibitem[{{Livne} {et~al.}(2005){Livne}, {Asida}, \&
  {H{\"o}flich}}]{livneetal05}
{Livne}, E., {Asida}, S.~M., \& {H{\"o}flich}, P. 2005, \apj, 632, 443

\bibitem[{{Marietta} {et~al.}(2000){Marietta}, {Burrows}, \&
  {Fryxell}}]{2000ApJS..128..615M}
{Marietta}, E., {Burrows}, A., \& {Fryxell}, B. 2000, \apjs, 128, 615

\bibitem[{{McClelland} {et~al.}(2010){McClelland}, {Garnavich}, {Galbany},
  {Miquel}, {Foley}, {Filippenko}, {Bassett}, {Wheeler}, {Goobar}, {Jha},
  {Sako}, {Frieman}, {Sollerman}, {Vinko}, \& {Schneider}}]{mcc+10}
{McClelland}, C.~M., {Garnavich}, P.~M., {Galbany}, L., {Miquel}, R., {Foley},
  R.~J., {Filippenko}, A.~V., {Bassett}, B., {Wheeler}, J.~C., {Goobar}, A.,
  {Jha}, S.~W., {Sako}, M., {Frieman}, J.~A., {Sollerman}, J., {Vinko}, J., \&
  {Schneider}, D.~P. 2010, \apj, 720, 704

\bibitem[{{Meakin} {et~al.}(2009){Meakin}, {Seitenzahl}, {Townsley}, {Jordan},
  {Truran}, \& {Lamb}}]{2009ApJ...693.1188M}
{Meakin}, C.~A., {Seitenzahl}, I., {Townsley}, D., {Jordan}, G.~C., {Truran},
  J., \& {Lamb}, D. 2009, \apj, 693, 1188

\bibitem[{{Niemeyer} \& {Woosley}(1997)}]{1997ApJ...475..740N}
{Niemeyer}, J.~C. \& {Woosley}, S.~E. 1997, \apj, 475, 740



\bibitem[{{Perlmutter} {et~al.}(1999){Perlmutter}, {Aldering}, {Goldhaber},
  {Knop}, {Nugent}, {Castro}, {Deustua}, {Fabbro}, {Goobar}, {Groom}, {Hook},
  {Kim}, {Kim}, {Lee}, {Nunes}, {Pain}, {Pennypacker}, {Quimby}, {Lidman},
  {Ellis}, {Irwin}, {McMahon}, {Ruiz-Lapuente}, {Walton}, {Schaefer}, {Boyle},
  {Filippenko}, {Matheson}, {Fruchter}, {Panagia}, {Newberg}, {Couch}, \& {The
  Supernova Cosmology Project}}]{1999ApJ...517..565P}
{Perlmutter}, S., {Aldering}, G., {Goldhaber}, G., {Knop}, R.~A., {Nugent}, P.,
  {Castro}, P.~G., {Deustua}, S., {Fabbro}, S., {Goobar}, A., {Groom}, D.~E.,
  {Hook}, I.~M., {Kim}, A.~G., {Kim}, M.~Y., {Lee}, J.~C., {Nunes}, N.~J.,
  {Pain}, R., {Pennypacker}, C.~R., {Quimby}, R., {Lidman}, C., {Ellis}, R.~S.,
  {Irwin}, M., {McMahon}, R.~G., {Ruiz-Lapuente}, P., {Walton}, N., {Schaefer},
  B., {Boyle}, B.~J., {Filippenko}, A.~V., {Matheson}, T., {Fruchter}, A.~S.,
  {Panagia}, N., {Newberg}, H.~J.~M., {Couch}, W.~J., \& {The Supernova
  Cosmology Project}. 1999, \apj, 517, 565

\bibitem[{{Phillips}(1993)}]{phillips93}
{Phillips}, M.~M. 1993, \apjl, 413, L105


\bibitem[{{Provencal} {et~al.}(1998){Provencal}, {Shipman}, {Hog}, \&
  {Thejll}}]{pro+98}
{Provencal}, J.~L., {Shipman}, H.~L., {Hog}, E., \& {Thejll}, P. 1998, \apj,
  494, 759

\bibitem[{{Pskovskii}(1977)}]{pskovskii77}
{Pskovskii}, I.~P. 1977, \sovast, 21, 675

\bibitem[{{Riess} {et~al.}(1998){Riess}, {Filippenko}, {Challis},
  {Clocchiatti}, {Diercks}, {Garnavich}, {Gilliland}, {Hogan}, {Jha},
  {Kirshner}, {Leibundgut}, {Phillips}, {Reiss}, {Schmidt}, {Schommer},
  {Smith}, {Spyromilio}, {Stubbs}, {Suntzeff}, \&
  {Tonry}}]{1998AJ....116.1009R}
{Riess}, A.~G., {Filippenko}, A.~V., {Challis}, P., {Clocchiatti}, A.,
  {Diercks}, A., {Garnavich}, P.~M., {Gilliland}, R.~L., {Hogan}, C.~J., {Jha},
  S., {Kirshner}, R.~P., {Leibundgut}, B., {Phillips}, M.~M., {Reiss}, D.,
  {Schmidt}, B.~P., {Schommer}, R.~A., {Smith}, R.~C., {Spyromilio}, J.,
  {Stubbs}, C., {Suntzeff}, N.~B., \& {Tonry}, J. 1998, \aj, 116, 1009

\bibitem[{{R{\"o}pke} \& {Hillebrandt}(2005)}]{2005A&A...431..635R}
{R{\"o}pke}, F.~K. \& {Hillebrandt}, W. 2005, \aap, 431, 635

\bibitem[{{R{\"o}pke} {et~al.}(2007){R{\"o}pke}, {Woosley}, \&
  {Hillebrandt}}]{2007ApJ...660.1344R}
{R{\"o}pke}, F.~K., {Woosley}, S.~E., \& {Hillebrandt}, W. 2007, \apj, 660,
  1344

\bibitem[{{Schmidt} {et~al.}(1998){Schmidt}, {Suntzeff}, {Phillips},
  {Schommer}, {Clocchiatti}, {Kirshner}, {Garnavich}, {Challis}, {Leibundgut},
  {Spyromilio}, {Riess}, {Filippenko}, {Hamuy}, {Smith}, {Hogan}, {Stubbs},
  {Diercks}, {Reiss}, {Gilliland}, {Tonry}, {Maza}, {Dressler}, {Walsh}, \&
  {Ciardullo}}]{schmidtetal98}
{Schmidt}, B.~P., {Suntzeff}, N.~B., {Phillips}, M.~M., {Schommer}, R.~A.,
  {Clocchiatti}, A., {Kirshner}, R.~P., {Garnavich}, P., {Challis}, P.,
  {Leibundgut}, B., {Spyromilio}, J., {Riess}, A.~G., {Filippenko}, A.~V.,
  {Hamuy}, M., {Smith}, R.~C., {Hogan}, C., {Stubbs}, C., {Diercks}, A.,
  {Reiss}, D., {Gilliland}, R., {Tonry}, J., {Maza}, J., {Dressler}, A.,
  {Walsh}, J., \& {Ciardullo}, R. 1998, \apj, 507, 46

\bibitem[{{Seitenzahl} {et~al.}(2009{\natexlab{a}}){Seitenzahl}, {Meakin},
  {Townsley}, {Lamb}, \& {Truran}}]{2009ApJ...696..515S}
{Seitenzahl}, I.~R., {Meakin}, C.~A., {Townsley}, D.~M., {Lamb}, D.~Q., \&
  {Truran}, J.~W. 2009{\natexlab{a}}, \apj, 696, 515

\bibitem[{{Seitenzahl} {et~al.}(2009{\natexlab{b}}){Seitenzahl}, {Townsley},
  {Peng}, \& {Truran}}]{2009ADNDT..95...96S}
{Seitenzahl}, I.~R., {Townsley}, D.~M., {Peng}, F., \& {Truran}, J.~W.
  2009{\natexlab{b}}, Atomic Data and Nuclear Data Tables, 95, 96

\bibitem[{{Timmes} \& {Swesty}(2000)}]{2000ApJS..126..501T}
{Timmes}, F.~X. \& {Swesty}, F.~D. 2000, \apjs, 126, 501

\bibitem[{{Townsley} {et~al.}(2007){Townsley}, {Calder}, {Asida}, {Seitenzahl},
  {Peng}, {Vladimirova}, {Lamb}, \& {Truran}}]{2007ApJ...668.1118T}
{Townsley}, D.~M., {Calder}, A.~C., {Asida}, S.~M., {Seitenzahl}, I.~R.,
  {Peng}, F., {Vladimirova}, N., {Lamb}, D.~Q., \& {Truran}, J.~W. 2007, \apj,
  668, 1118

\bibitem[{{Valenti} {et~al.}(2009){Valenti}, {Pastorello}, {Cappellaro},
  {Benetti}, {Mazzali}, {Manteca}, {Taubenberger}, {Elias-Rosa}, {Ferrando},
  {Harutyunyan}, {Hentunen}, {Nissinen}, {Pian}, {Turatto}, {Zampieri}, \&
  {Smartt}}]{val+09}
{Valenti}, S., {Pastorello}, A., {Cappellaro}, E., {Benetti}, S., {Mazzali},
  P., {Manteca}, J., {Taubenberger}, S., {Elias-Rosa}, N., {Ferrando}, R.,
  {Harutyunyan}, A., {Hentunen}, V.-P., {Nissinen}, M., {Pian}, E., {Turatto},
  M., {Zampieri}, L., \& {Smartt}, S.~J. 2009, ArXiv, 0901.2074



\end{thebibliography}




\begin{deluxetable*}{ccccccccccccccc}
\tabletypesize{\scriptsize}
\tablecaption{Simulations Properties\label{tab:sims}}
\tablewidth{0pt}
\tablehead{
\colhead{sim} &
\colhead{$\Delta x$\tablenotemark{a}} &
\colhead{n$_{\rm ign}$\tablenotemark{b}} &
\colhead{r$_{\rm ign}$\tablenotemark{c}} &
\colhead{z$_{\rm ign}$\tablenotemark{d}} &
\colhead{E$_{\rm nuc}$\tablenotemark{e}} &
\colhead{C-O$_{\rm B}$\tablenotemark{f}} &
\colhead{IME$_{\rm B}$} &
\colhead{IGE$_{\rm B}$} &
\colhead{v$_{\rm B}$\tablenotemark{g}} &
\colhead{C-O$_{\rm E}$\tablenotemark{h}} &
\colhead{IME$_{\rm E}$} &
\colhead{IGE$_{\rm E}$}  &
\colhead{E$_{\rm kin,E}$\tablenotemark{i}} &
\colhead{v$_{\rm ave,E}$\tablenotemark{j}} \\
\colhead{name} &
\colhead{(km)} &
\colhead{} &
\colhead{(km)} &
\colhead{(km)} &
\colhead{(E$_{bin}$)} &
\colhead{(\msolar)} &
\colhead{(\msolar)} &
\colhead{(\msolar)} &
\colhead{(km s$^{-1}$)} &
\colhead{(\msolar)} &
\colhead{(\msolar)} &
\colhead{(\msolar)} &
\colhead{(ergs $\cdot 10^{50}$)} &
\colhead{(km s$^{-1}$)}
}
\startdata
2D70 &       4       &      4      &   64.0        &   70.0       & 0.89 & 0.93  & 0.07  & 0.13  & 119    & 0.13  & 0.03  & 0.07  & 0.32  &  3,730 \\
3D48 &       8       &     63      &   128.0       &   48.0       & 1.06 & 0.84  & 0.06  & 0.09  & 351    & 0.16  & 0.05  & 0.16  & 0.90  &  4,932 \\
3D38 &       8       &     63      &   128.0       &   38.0       & 1.30 & 0.68  & 0.05  & 0.05  & 411    & 0.27  & 0.09  & 0.24  & 1.6   &  5,139 \\
3D28 &       8       &     63      &   128.0       &   28.0       & 1.51 & 0.49  & 0.03  & 0.04  & 549    & 0.41  & 0.13  & 0.30  & 2.3   &  5,229 \\
3D18 &       8       &     63      &   128.0       &   18.0       & 1.67 & 0.26  & 0.02  & 0.02  & 483    & 0.58  & 0.17  & 0.34  & 2.9   &  5,193 \\
\enddata
\tablenotetext{a}{Maximum spatial resolution.}
\tablenotetext{b}{Number of ignition points.}
\tablenotetext{c}{Radius of the spherical volume containing the ignition points.}
\tablenotetext{d}{Z coordinate of the origin of the spherical volume containing the ignition points.}
\tablenotetext{e}{Energy released during the deflagration phase divided by the binding energy of the star ($E_{bin} = 4.5\times 10 ^{50}$ ergs).}
\tablenotetext{f}{``B'' refers to gravitationally bound material.}
\tablenotetext{g}{Velocity of the gravitationally bound material .}
\tablenotetext{h}{``E'' refers to material that will escape the system.}
\tablenotetext{i}{Average kinetic energy of the escaping material.}
\tablenotetext{j}{Mass weighted velocity of escaping material.}
\end{deluxetable*}

\end{document}